\documentclass[onecolumn,11pt,showpacs,preprintnumbers,amsmath,amssymb]{revtex4}

\usepackage{graphicx}
\usepackage{dcolumn}
\usepackage{bm}
\usepackage{subfigure}
\usepackage{marvosym}
\usepackage{textcomp}
\usepackage{threeparttable}
\usepackage{booktabs}
\usepackage{setspace}
\usepackage{latexsym}
\usepackage{color}
\renewcommand{\TPTtagStyle}%
{\normalsize\textit}

\begin{document}

\title{Light Probes in a Strongly Coupled Anisotropic Plasma}
\author{Berndt M\"uller and Di-Lun Yang}
\affiliation{Department of Physics, Duke University, Durham, North Carolina 27708, USA}%

\begin{abstract}
We employ the gauge/gravity duality to study the jet quenching of light probes traversing a static yet anisotropic strongly coupled $\mathcal{N}=4$ super Yang-Mills plasma. We compute the stopping distance of an image jet induced by a massless source field, which is characterized by a massless particle falling along the null geodesic in the WKB approximation, in an anisotropic dual geometry introduced by Mateos and Trancancelli. At mid and large anisotropic regimes, the stopping distances of a probe traveling in the anisotropic plasma along various orientations are suppressed compared to those in an isotropic plasma especially along the longitudinal direction at equal temperature. However, when fixing the entropy density, the anisotropic values of stopping distances near the transverse directions slightly surpass the isotropic values. In general, the jet quenching of light probes is increased by the anisotropic effect in a strongly coupled and  equilibrium plasma.  

\end{abstract}
\maketitle
\section{\label{sec:level1}Introduction}
The AdS/CFT correspondence(or gauge/gravity duality) \cite{Maldacena:1997re,Gubser:1998bc,Witten:1998qj,Witten:1998zw,Aharony:1999ti}  is a duality between the $\mathcal{N}=4$ $SU(N_c)$ super Yang-Mills theory(SYM) and the type IIB supergravity in the Anti-de Sitter space $AdS_5\times S^5$ at the large t'Hooft coupling and large $N_c$ limit, which has been widely utilized to tackle strongly coupled systems. Various findings from the quark gluon plasma(QGP) generated in the heavy ion collisions suggest that QGP behaves as a nearly perfect fluid \cite{Shuryak:2003xe,Shuryak:2004cy,Adcox:2004mh,Adams:2005dq,Luzum:2009sb,Song:2010mg,Qiu:2011hf,Schenke:2011bn,Gale:2012rq}, 
which involves non-perturbative physics suitable for the application of the AdS/CFT duality. Although the $\mathcal{N}=4$ SYM is conformal, which differs from QCD in many properties, it may reveal some of qualitative features of QCD in the strongly coupled regime. One of the renowned works is the universal lower bound of the ratio of shear viscosity over entropy density found in the holographic model \cite{Policastro:2001yc,Kovtun:2004de}.  

Although the ideal hydrodynamics successfully describes many properties of QGP based on experimental data, further studies propose that the medium should be anisotropic in the early times during the formation of QGP \cite{Florkowski:2008ag,Florkowski:2009sw,Ryblewski:2010tn,Florkowski:2010cf,Martinez:2010sc,Martinez:2010sd,Ryblewski:2010ch,Ryblewski:2011aq,Strickland:2011mw,Strickland:2011aa,Martinez:2012tu,Florkowski:2012as}. Within the time period $\tau_0\sim Q_s^{-1}\leq\tau\leq\tau_{hydro}$, where $Q_s$ represents the saturation scale of colliding nuclei, the pressure of the medium along the transverse direction (perpendicular to the beam axis) may surpass the pressure along the longitudinal direction (parallel to the beam axis). Moreover, the recent study on the thermalization of a strongly coupled plasma via AdS/CFT correspondence suggests an anisotropic medium even in the late times \cite{Heller:2011ju,Heller:2012je}. The authors examined a wide variety of initial conditions and found all setups lead to anisotropic hydrodynamics after the thermalization. Therefore, at least withing a certain time period, QGP could be dictated by anisotropic hydrodynamics with unequal pressures along distinct directions. Despite the spacial expansion of QGP with respect to time, the investigation of a static yet anisotropic plasma may shed some light on the physics in early times. In \cite{Janik:2008tc} and \cite{Mateos:2011ix,Mateos:2011tv}, the dual geometries of such a strongly coupled and static plasma with anisotropy are derived. The former is originated from solely the anisotropic energy stress tensor, while the spacetime has a naked singularity. The latter is generated from a five-dimensional effective action with a dilaton and an axion field linearly depending on an anisotropic factor, which results in a regular geometry. The further study of a static and anisotropic plasma with finite chemical potentials can be found in \cite{Gahramanov:2012wz}.    

The energy loss of hard probes in such strongly-coupled anisotropic plasma described by holographic models has been investigated recently.
The energy loss of heavy probes in the gravity dual can be characterized by the drag force obtained from a trailing string moving in the dual geometry \cite{Gubser:2006qh,Herzog:2006gh} or the jet quenching parameter extracted from a lightcone Wilson loop probing the infrared scale \cite{Liu:2006ug,Liu:2006he}. The jet quenching parameter, drag force as well as the heavy quark potential are computed in holographic duals \cite{Giataganas:2012zy,Rebhan:2012bw,Chernicoff:2012gu,Chernicoff:2012iq}. In addition, the energy loss of orbiting quarks and quarkonium dissociation in an anisotropic holographic dual also have been explored in  \cite{Fadafan:2012qu} and \cite{Chernicoff:2012bu}, respectively. It is found in \cite{Giataganas:2012zy,Chernicoff:2012iq} that the drag force in the anisotropic plasma can be smaller or larger than the isotropic case depending on different velocities and orientations of the probes. On the other hand, the anisotropic value of the jet quenching parameter may be enhanced or reduced when comparing to the isotropic value at equal temperature or at equal entropy density \cite{Chernicoff:2012gu,Rebhan:2012bw}.    
These studies focus on the jet quenching of heavy probes in the gravity dual, whereas the anisotropic effect on light probes
has not yet been investigated. In the isotropic case, the jet quenching of light probes can be characterized by a maximum stopping distance of a massless particle falling along the null geodesic in the dual geometry based on various approaches \cite{Gubser:2008as,Chesler:2008uy,Arnold:2010ir,Arnold:2011qi,Arnold:2012uc}. We will thus carry out the computation of the stopping distance in the anisotropic dual geometry introduced from \cite{Mateos:2011ix,Mateos:2011tv}. 

In Section II, we will compute the stopping distance of a light probe at small anisotropy or equivalently at high temperature by tracking a massless particle falling in the anisotropic geometry up to the leading-order expansion of the ratio of anisotropic factor over temperature. We then present the numerical results of stopping distances at mid and large anisotropy in Section III and make a brief summary in the final section.     

\section{\label{sec:level1}Stopping Distances at Small Anisotropy}
In this section, we will investigate the maximum stopping distance of a light probe in an anisotropic plasma at the high-temperature or the small-anisotropy regime. In \cite{Mateos:2011ix, Mateos:2011tv}, Mateos and Trancanelli(MT) introduced a five-dimensional dilaton-axion action, which leads to an anisotropic solution of the spacetime metric. In the string frame, 
the solution is given by
\begin{eqnarray}\label{MT}
ds^2=\frac{1}{z^2}\left(-\mathcal{F}(z)\mathcal{B}(z)dt^2+dx_T^2+\mathcal{H}(z)dx_L^2+\frac{dz^2}{\mathcal{F}(z)}\right),
\end{eqnarray}
where $dx_T^2=(dx^1)^2+(dx^2)^2$ and $dx_L^2=(dx^3)^2$ represent the transverse direction and the longitudinal direction, respective. The metric is generated by a dilaton $\phi(z)$ and an axion $\chi(z)=ax_L$ encoding the anisotropic factor $a$, where $a$ represents the number density of D7-branes along the longitudinal direction as a magnetic source of the axion. In heavy ion collisions, the longitudinal direction could be regarded as the beam direction. The nonzero anisotropic factor gives rise to the pressure difference between the directions parallel and perpendicular to beams. Thus, the dual geometry may mimic a boost invariant medium along the beam direction without expansion.    

In the high-temperature or the small-anisotropy regime $(T\gg a)$, the pressure anisotropy is small. In this regime, the anisotropic factor $a$ in the strongly coupled scenario can be related to a parameter $\xi$ introduced in the weakly coupled approach \cite{Romatschke:2003ms,Romatschke:2004jh}, which is define
as 
\begin{eqnarray}
\xi=\frac{\langle p_T^2\rangle}{2\langle p_L^2\rangle}-1, 
\end{eqnarray}
where $p_T$ and $p_L$ denote the magnitudes of momenta along transverse and longitudinal directions, respectively. 
The parameter $\xi$ characterizes the momentum anisotropy in a weakly coupled and anisotropic plasma. It was found in \cite{Giataganas:2012zy} that
\begin{eqnarray}
\xi\approx\frac{5a^2}{8\pi^2T^2}~\mbox{for}~0<\xi\ll 1
\end{eqnarray}
by comparing the pressure differences in two approaches. When $T\gg a$,
the analytic expression of the metric up to the leading order can be found,
\begin{eqnarray}\label{comp}
\nonumber\mathcal{F}(z)&=&1-\frac{z^4}{z_h^4}+a^2\mathcal{F}_2(z)+\mathcal{O}(a^4),\\
\nonumber\mathcal{B}(z)&=&1+a^2\mathcal{B}_2(z)+\mathcal{O}(a^4),\\
\mathcal{H}(z)&=&e^{-\phi(z)},\mbox{}\phi(z)=a^2\phi_2(z)+\mathcal{O}(a^4),
\end{eqnarray}
where $z_h$ is the event horizon such that $\mathcal{F}(z_h)=0$. From (\ref{comp}), the MT metric in (\ref{MT}) will reduce to the AdS-Schwarzschild metric when $a\rightarrow 0$. The explicit expression of the leading-order anisotropic terms are 
\begin{eqnarray}
\nonumber\mathcal{F}_2(z)&=&\frac{1}{24z_h^2}\left[8z^2(z_h^2-z^2)-10z^4\mbox{log}2+(3z_h^4+7z^4)\mbox{log}\left(1+\frac{z^2}{z_h^2}\right)\right],\\
\nonumber\mathcal{B}_2(z)&=&-\frac{z_h^2}{24}\left[\frac{10}{z_h^2+z^2}+\mbox{log}\left(1+\frac{z^2}{z_h^2}\right)\right],\\
\phi_2(z)&=&-\frac{z_h^2}{4}\mbox{log}\left(1+\frac{z^2}{z_h^2}\right).
\end{eqnarray} 
The temperature at the leading order is defined as
\begin{eqnarray}
T=-\frac{\partial_z(\mathcal F\sqrt{\mathcal{B}})|_{z=z_h}}{4\pi}=\frac{1}{\pi z_h}+a^2z_h\frac{5\mbox{log}2-2}{48\pi}+\mathcal{O}(a^4).
\end{eqnarray}
Conversely, the horizon can be written as
\begin{eqnarray}\label{zhorizon}
z_h=\frac{1}{\pi T}+a^2\frac{5\mbox{log}2-2}{48\pi^3T^3}+\mathcal{O}(a^4).
\end{eqnarray}

Now, we may study the jet quenching of a light probe in the anisotropic plasma by computing the stopping distance of a massless particle moving along the null geodesic in the MT metric. We will follow the approach in \cite{Hatta:2008tx,Arnold:2010ir,Arnold:2011qi}, where an R-charged current is generated by a massless gauge field in the gravity dual. The induced current may be regarded as an energetic jet traversing the medium. When the wave packet of the massless field falls into the horizon of the dual geometry, the image jet on the boundary dissipates and thermalizes in the medium. The stopping distance is thus define as the distance for a jet traversing the medium before it thermalizes. In the WKB approximation, we assume that the wave packet of the gauge field in the gravity dual highly localized in the momentum space. We thus factorize the wave function of the gauge field as
\begin{eqnarray}
A_j(t,z)=\exp\left[\frac{i}{\hbar}\left(q_kx^k+\int dz q_z\right)\right]\tilde{A}_j(t,z),
\end{eqnarray} 
where $q_z$ denotes the momentum along the bulk direction and $j,k=0,1,2,3$ represent four-dimensional spacetime coordinates and $q_{k}$ denotes the four-momentum, which is conserved as the metric preserves the translational symmetry along the four-dimensional spacetime. Here $\tilde{A}_i(t,z)$ is slow-varying with respect to $t$ and $z$. In the classical limit($\hbar\rightarrow 0$), the equation of motion of the wave packet will reduce to a null geodesic in the dual geometry, which takes the form \cite{Arnold:2011qi},
\begin{eqnarray}\label{geodesic}
\frac{dx^i}{dz}=\sqrt{g_{zz}}\frac{g^{ij}q_{j}}{(-q_{k}q_{l}g^{kl})^{1/2}}.
\end{eqnarray}
Thus, the wave packet can be approximated as a massless particle and the null geodesic will lead to a maximum stopping distance for an image jet on the boundary in the classical limit. In \cite{Chesler:2008uy}, the bulk governed by the AdS-Schwarzschild geometry is filled with a D7 brane, where the backreaction of the D7 brane to the bulk geometry is ignored. A string falling in the bulk then induces a flavor-current on the boundary, which can be regarded as a light quark traversing the medium. Notice that the flavor D7 brane here is different from the D7 brane in MT model as the source of anisotropy. When the tip of the string falls into the horizon, the flavor current fully diffuses on the boundary, which corresponds to the thermalization of a light quark in the medium. In this scenario, a null geodesic will result in the maximum stopping distance for the light quark, which is similar to the previous setup for an R-charged current. 
 
By using (\ref{geodesic}), the stopping distance in an isotropic medium governed by the AdS-Schwarzschild spacetime is given by
\begin{eqnarray}\label{xstop}
x_s=\int^{z_H}_0\frac{dz}{\left(\frac{\omega^2}{|\vec{q}|^2}-F(z)\right)^{1/2}},
\end{eqnarray}
where $F(z)=1-z^4/z_H^4$ and $z_H$ denotes the position of the event horizon. Here $\omega$ and $\vec{q}$ represent the energy and the spacial momentum of the particle, respectively. Analogously, by employing (\ref{MT}) and (\ref{geodesic}), we can compute the stopping distances of the probes traveling along the transverse and the longitudinal directions in the anisotropic medium,
\begin{eqnarray}\label{xT}
x_T&=&\int^{z_h}_0\frac{dz}{\left(\frac{1}{\mathcal{B}}\frac{\omega^2}{|\vec{q}|^2}-\mathcal{F}\right)^{1/2}},\\
\label{xL}x_L&=&\int^{z_h}_0\frac{dz}{\mathcal{H}\left(\frac{1}{\mathcal{B}}\frac{\omega^2}{|\vec{q}|^2}-\frac{\mathcal{F}}{\mathcal{H}}\right)^{1/2}},
\end{eqnarray} 
where we assume that the particle carries the spatial momentum solely along one of the transverse direction in (\ref{xT}) and solely along the longitudinal direction in (\ref{xL}). In this computation, the null geodesic in (\ref{geodesic}) remains unchanged even when we use the Einstein frame.

\begin{figure}[h]
{\includegraphics[width=7.5cm,height=5cm,clip]{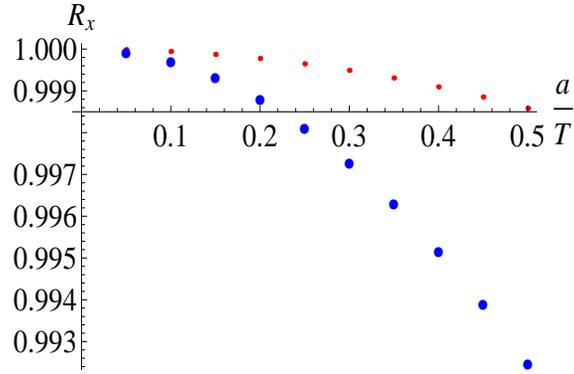}}
\caption{$R_x=x_{aniso}/x_{iso}$ represents the ratio of the stopping distances in the MT geometry with anisotropy to without anisotropy, where $x_{aniso}=x_T$ for the red points and $x_{aniso}=x_L$ for the large blue points. Here we take $|\vec{q}|=0.99\omega$ and fix the temperature of media.}
\label{MTplot}
\vspace{3mm}
\end{figure}

\begin{figure}[h]
\begin{minipage}{7cm}
\begin{center}
{\includegraphics[width=7.5cm,height=5cm,clip]{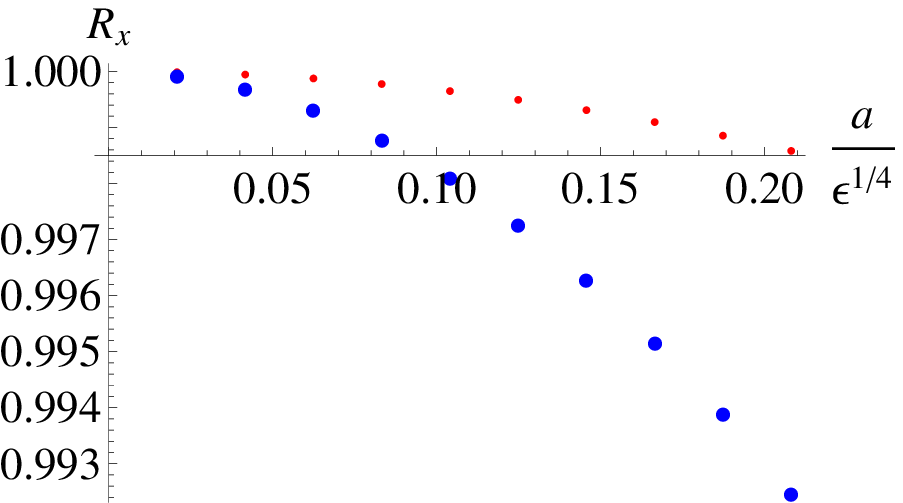}}
\caption{$R_x=x_{aniso}/x_{iso}$ represents the ratio of the stopping distances in the MT geometry with anisotropy to without anisotropy, where $x_{aniso}=x_T$ for the red points and $x_{aniso}=x_L$ for the large blue points. Here we take $N_c=3$, $|\vec{q}|=0.99\omega$ and fix the energy density of media.}\label{fixE}
\end{center}
\end{minipage}
\hspace {1cm}
\begin{minipage}{7cm}
\begin{center}
{\includegraphics[width=7.5cm,height=5cm,clip]{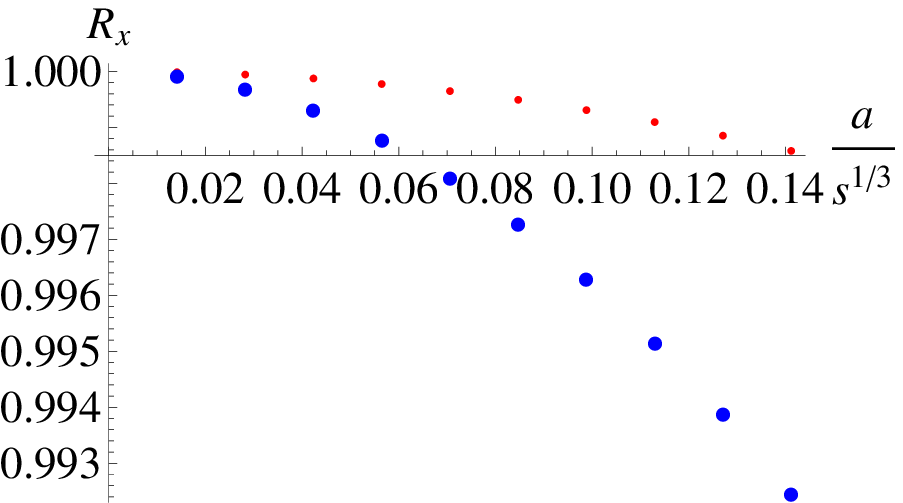}}
\caption{$R_x=x_{aniso}/x_{iso}$ represents the ratio of the stopping distances in the MT geometry with anisotropy to without anisotropy, where $x_{aniso}=x_T$ for the red points and $x_{aniso}=x_L$ for the large blue points. Here we take $N_c=3$, $|\vec{q}|=0.99\omega$ and fix the entropy density of media.}\label{fixs}
\end{center}
\end{minipage}
\end{figure}

To compare the stopping distance in the media with and without the anisotropic effect, we have to fix a proper physical parameter. In the following computation, we will fix the temperature, energy density, and entropy density, respectively. The recent lattice simulation for the $SU(N_c)$ plasma at finite temperature within the range of RHIC and LHC has shown that the equilibrium thermodynamic properties have mild dependence of $N_c$ \cite{Panero:2009tv}, which supports the validity of the study of QCD based on large $N_c$ models. In general, only when a particular observable obtained from the lattice calculation matches that found by AdS/CFT, then the lattice findings can be used for the extrapolation to the small-$N_c$ limit. However, for an observable which does not depend on $N_c$ explicitly, such as the ratio of stopping distances for our concern in the paper, the results in the large-$N_c$ and in the small-$N_c$ limits may share same features qualitatively. When fixing the energy density and entropy density, we will always take $N_c=3$ in analogy to QCD, while the choice of $N_c$ will not affect our qualitative results in this paper. From (\ref{xT}) and (\ref{xL}), we see that the parameter-dependence of the stopping distance is encoded by the position of the horizon in terms of the physical parameter we fix. By inserting (\ref{zhorizon}) into (\ref{xT}) and (\ref{xL}), we can compute the stopping distance by fixing the temperature of the medium.  
The stopping distances for different values of the anisotropy factor $a$ in units of temperature are illustrated in Fig.\ref{MTplot}. In \cite{Mateos:2011tv}, the energy density and entropy density up to leading order in $a^2$ are given by
\begin{eqnarray}\label{Density}\nonumber
\epsilon&=&\frac{3N_c^2\pi^2T^4}{8}+a^2\frac{N_c^2T^2}{32}+\mathcal{O}(a^4),\\
s&=&\frac{N_c^2\pi^2T^3}{2}+a^2\frac{N_c^2T}{16}+\mathcal{O}(a^4).
\end{eqnarray}
Combining (\ref{zhorizon}) and (\ref{Density}), we rewrite the position of the horizon in terms of the energy density or the entropy density,
\begin{eqnarray}\label{zhorizond}\nonumber
z_h&=&\left(\frac{3N_c^2}{8\pi^2}\right)^{1/4}\epsilon^{-1/4}+a^2\frac{5\log 2-1}{128\pi^2}\left(\frac{8\pi^2}{3N_c^2}\right)^{1/4}N_c^2\epsilon^{-3/4}+\mathcal{O}(a^4),\\
z_h&=&\left(\frac{\pi^2N_c^2}{2}\right)^{1/3}\frac{s^{-1/3}}{\pi}+a^2\frac{5\log 2}{96\pi}N_c^2s^{-1}+\mathcal{O}(a^4).
\end{eqnarray}  
By utilizing (\ref{xT}), (\ref{xL}), and (\ref{zhorizond}), we are now able to compute the stopping distances for fixed energy density or fixed entropy density. The results are shown in Fig.\ref{fixE} and Fig.\ref{fixs}.  
We find that the nonzero anisotropic factor leads to smaller stopping distances in both the transverse direction and the longitudinal direction, which indicates stronger jet quenching of light probes traveling through the anisotropic medium. The quenching is more pronounced along the longitudinal direction. Although this effect is rather small, it is not surprising since the momentum anisotropy for $a\leq T$ is rather small as shown in \cite{Giataganas:2012zy}. In contrast to light probes, the jet quenching of heavy probes is also weakly enhanced at small anisotropy or at high temperature. As shown in \cite{Giataganas:2012zy}, only slightly greater jet quenching parameters and drag forces of heavy probes moving beyond the critical velocity are found in the MT geometry at $a/T=0.3$. 

We now analyze the relation between the energy density and the stopping distance in the anisotropic plasma in more detail. In the AdS-Schwarzschild spacetime, the energy density is characterized by the temperature, $\epsilon=3\pi^2N_c^2T^4/8$. By using (\ref{xstop}), the stopping distance can be rewritten as
\begin{eqnarray}
x_s=\epsilon^{-1/4}\left(\frac{3N_c^2}{8\pi^2}\right)\int^{1}_0\frac{dr}{\left(-\frac{q^2}{|\vec{q}|^2}+r^4\right)^{1/2}}
=\epsilon^{-1/4}A_0(N_c,\omega,|\vec{q}|),
\end{eqnarray}
where $r=z/z_H$ and $q^2=-\omega^2+|\vec{q}|^2$ in the integral and $A_0(N_c,\omega,|\vec{q}|)$ is a dimensionless factor. We see that the stopping distance of the hard probe decreases when the energy density is increased. This is analogous to the weakly-coupled plasma where the jet quenching is enhanced for increasing energy density \cite{Baier:2002tc}. When including the anisotropic effect, the transverse stopping distance to the order $a^2$ becomes
\begin{eqnarray}\nonumber\label{xTinE}
x_T&=&\int^1_0dr\frac{z_h^0}{\left(-\frac{q^2}{|\vec{q}|^2}+r^4\right)^{1/2}}
\left[1+\frac{a^2}{2\left(-\frac{q^2}{|\vec{q}|^2}+r^4\right)}\left(\mathcal{B}_2\frac{\omega^2}{|\vec{q}|^2}+\mathcal{F}_2\right)+a^2\delta z_h^0\right]\\
&=&\epsilon_a^{-1/4}A_0(N_c,\omega,|\vec{q}|)+a^2\epsilon_a^{-3/4}A_1^T(N_c,\omega,|\vec{q}|),
\end{eqnarray}
where $z_h^0$ and $\delta z_h^0$ can be read off from (\ref{zhorizond}). Here $\epsilon_a$ is the energy density shown in (\ref{Density}) and we may take $\epsilon_a=\epsilon$ for comparison. For the hard probe with small virtuality, the dimensionless factor $A_1^T(N_c,\omega,|\vec{q}|)$ is negative; hence the suppression of the stopping distance led by the first-order anisotropic correction is reduced when the energy density is increased. The longitudinal stopping distance $x_L$ can be written in the same form as $x_T$ by substituting $A_1^T(N_c,\omega,|\vec{q}|)$ with a different numerical factor $A_1^L(N_c,\omega,|\vec{q}|)$, where $A_1^L(N_c,\omega,|\vec{q}|)$ is also negative at small virtuality. The similar scenario for fixed entropy density could be found by following the same approach.     

In general, in the high-temperature or small-anisotropy limit, the anisotropic values of stopping distances are slightly smaller than the isotropic values by fixing one of physical parameters such as temperature, energy density, or entropy density. The jet quenching along the longitudinal direction is particularly enhanced, although the enhancement is rather small.    

\section{\label{sec:level1}Stopping Distances at Mid and Large Anisotropy}
In the mid-anisotropy or the large-anisotropy regimes, the longitudinal pressure surpasses the transverse pressure and the pressure inequality becomes substantial \cite{Mateos:2011tv}, which may not be similar to the feature of QGP. In anisotropic hydrodynamics, the pressure difference can be considerably large in early times \cite{Martinez:2010sc,Martinez:2012tu}. However, the longitudinal pressure should be always suppressed by the transverse pressure, which is qualitatively opposite to the scenario described by the MT model beyond small anisotropy. Despite the unrealistic directions of anisotropy, it may be heuristic to investigate the effect of the medium with strong anisotropy on jet quenching of light probes. 

\begin{figure}[h]
\begin{minipage}{7cm}
\begin{center}
{\includegraphics[width=7.5cm,height=5cm,clip]{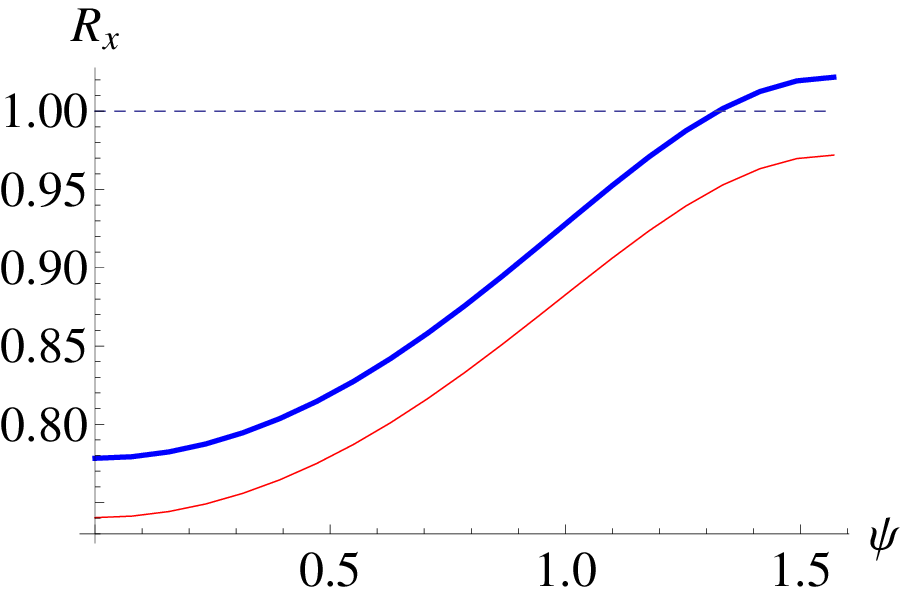}}
\caption{The red and thick blue curves represent the ratios $R_x=x_{aniso}/x_{iso}$ at mid anisotropy at equal temperature and at equal entropy density, respectively. Here we take $|\vec{q}|=0.99\omega$, $z_h=1$, and $a/T\approx 4.4$ or equivalently $a/s^{1/3}\approx 1.2$ for $N_c=3$.}\label{xrpsi1}
\end{center}
\end{minipage}
\hspace {1cm}
\begin{minipage}{7cm}
\begin{center}
{\includegraphics[width=7.5cm,height=5cm,clip]{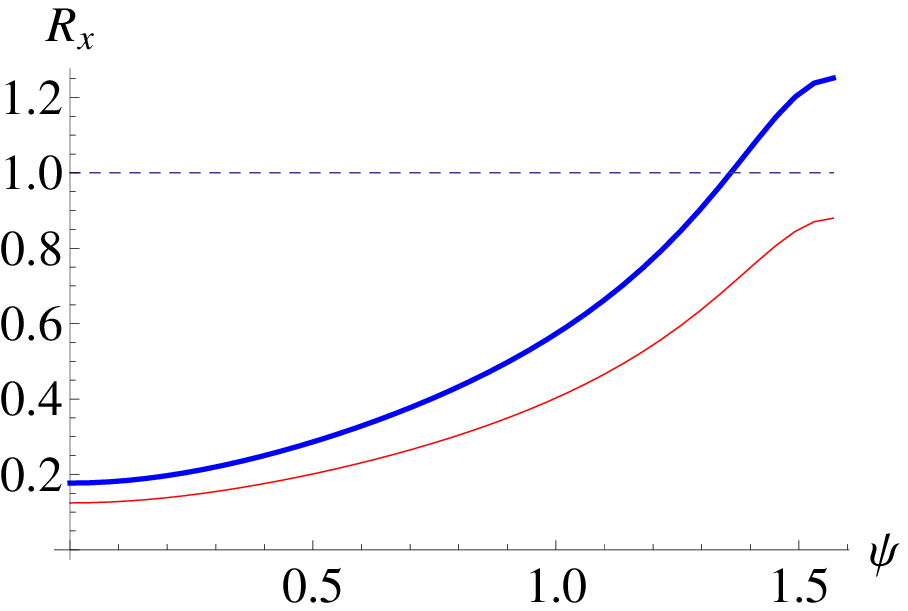}}
\caption{The red and thick blue curves represent the ratios $R_x=x_{aniso}/x_{iso}$ at large anisotropy at equal temperature and at equal entropy density, respectively. Here we take $|\vec{q}|=0.99\omega$, $z_h=1$, and $a/T\approx 86$ or equivalently $a/s^{1/3}\approx 17$ for $N_c=3$.}\label{xrpsi3}
\end{center}
\end{minipage}
\end{figure}

To study the jet quenching in the mid-anisotropy or the large-anisotropy regimes, we have to employ the numerical solution of the MT metric. As shown in \cite{Mateos:2011tv}, the spacetime metric is derived from solving the Einstein equations and the dilaton equation with the only restriction $\mathcal{F}(z_h)=0$ by taking the following expansion,
\begin{eqnarray}
\tilde{\phi}(z)=\tilde{\phi}_h+\sum_{n\geq 1}\tilde{\phi}_n(z-z_h)^n,
\end{eqnarray}  
where $\tilde{\phi}(z)=\phi(z)+\log a^{4/7}$. By fixing $\phi(0)=0$, the anisotropic factor will be determined by $a=e^{\frac{7}{4}\tilde{\phi}(0)}$. The numerical result of the metric should only depend on the values of $\tilde{\phi}_h$ and $z_h$. The full expressions of the temperature and entropy density are 
\begin{eqnarray}\nonumber
T&=&\sqrt{\mathcal{B}(z_h)}\frac{e^{\frac{1}{2}(\tilde{\phi}_{b}-\tilde{\phi}_h)}}{16\pi z_h}(16+z_h^2e^{\frac{7}{2}\tilde{\phi}_h}),\\
s&=&\frac{N_c^2a^{\frac{5}{7}}e^{-\frac{5}{4}\tilde{\phi}_h}}{2\pi z_h^3},
\end{eqnarray} 
where $\tilde{\phi}_b=\tilde{\phi}(0)$. Here we are interested in the stopping distance along an arbitrary direction. Due to the rotational symmetry in the transverse directions, we may set the four-momentum of the massless particle as $q_i=(-\omega,|\vec{q}|\sin\psi,0,|\vec{q}|\cos\psi)$, where $\psi$ denotes the polar angle in the $x^1-x^3$ plane with respect to the longitudinal direction $x^3$. The probe thus travels along the longitudinal and transverse directions for $\psi=0$ and $\psi=\pi/2$, respectively. Now the stopping distance acquired from (\ref{geodesic}) is given by $x_{aniso}=\sqrt{x_{Ts}^2+x_{Ls}^2}$, where
\begin{eqnarray}\nonumber
x_{Ts}&=&\int^{z_h}_0dz\frac{\sin\psi}{\left(\frac{1}{\mathcal{B}}\frac{\omega^2}{|\vec{q}|^2}-\frac{\mathcal{F}}{\mathcal{H}}(\cos^2\psi+\mathcal{H}\sin^2\psi)\right)^{1/2}},\\
x_{Ls}&=&\int^{z_h}_0dz\frac{\cos\psi}{\mathcal{H}\left(\frac{1}{\mathcal{B}}\frac{\omega^2}{|\vec{q}|^2}-\frac{\mathcal{F}}{\mathcal{H}}(\cos^2\psi+\mathcal{H}\sin^2\psi)\right)^{1/2}}.
\end{eqnarray} 
By inserting the numerical solutions of the spacetime metric into the equation above and carrying out the integrations, the stopping distances at mid and large anisotropy in comparison with those in the isotropic case are shown in Fig.\ref{xrpsi1} and Fig.\ref{xrpsi3}. When fixing the temperature, the stopping distances in both mid and large anisotropy are smaller compared to the isotropic results. As shown in both figures, when $\psi$ decreases, the suppression of the stopping distance becomes more robust, which suggests stronger jet quenching along the longitudinal direction. In contrast, at equal entropy density, the enhanced jet quenching in the anisotropic medium becomes less prominent and the stopping distances of probes moving close to the transverse directions even exceed the stopping distances in the isotropic case. Overall, the jet quenching of light probes is enhanced when turning up the anisotropic effect except for the probe moving along the transverse direction.

After finding the results at mid and large anisotropy, we may make a comparison with the influence of the anisotropic effect on the jet quenching of heavy probes. In the studies of the drag force in MT metric \cite{Giataganas:2012zy,Chernicoff:2012iq}, the longitudinal drag is as well enhanced by anisotropy at equal temperature or at equal entropy density . Also, at mid or large anisotropy, the enhancement of the drag force for the probe traveling more parallel to the transverse direction diminishes. When the magnitude of the probe velocity is smaller than a critical value, the transverse drag could be smaller than the isotropic drag at equal temperature or equal entropy density. Despite the velocity dependence, the angular dependence of the anisotropic drag is qualitatively analogous to the scenario of the anisotropic stopping distance we find. For the jet quenching parameter computed from lightcone Wilson loops, the enhancement or the suppression due to anisotropy are more subtle, which depends on both the direction of moving quarks and the direction of momentum broadening \cite{Chernicoff:2012gu,Rebhan:2012bw}. Nevertheless, since the momentum broadening is attributed to collisions between the heavy quark and thermal partons in the medium, this effect should be suppressed compared to the radiation energy loss in the case of light probes. As a result, we may not anticipate that the anisotropy effect on stopping distances of light probes shares the same features with the jet quenching parameters.     
 
\section{\label{sec:level1}Summary}  
In this paper, we have calculated the maximum stopping distance of an energetic jet traveling in a strongly coupled anisotropic plasma by analyzing the null geodesic of a massless particle falling in the dual geometry. We carried out the investigation from low anisotropy to large anisotropy. At small anisotropy, the stopping distances slightly decrease in comparison with the isotropic case for fixed temperature, energy density, and entropy density, respectively. At mid or large anisotropy, we found that the anisotropic stopping distances are generally smaller than the isotropic ones at equal temperature or equal entropy density especially along the longitudinal direction. However, along the transverse direction, the suppression of the stopping distance becomes less prominent at equal temperature. When fixing the entropy density, the transverse stopping distance is even larger than the isotropic one. In the end, we discussed the similarity between the anisotropic stopping distances of light probes we found and the anisotropic drag forces of heavy probes in previous literature.

\bigbreak\bigskip\bigskip\centerline{{\bf Acknowledgements}}\nobreak
\bigskip
The authors thank P. Arnold and D. Vaman for useful discussions. This work was supported by DOE grants DE-FG02-05ER41367 and DE-SC0005396. 


\begin{thebibliography}{54}
\expandafter\ifx\csname natexlab\endcsname\relax\def\natexlab#1{#1}\fi
\expandafter\ifx\csname bibnamefont\endcsname\relax
  \def\bibnamefont#1{#1}\fi
\expandafter\ifx\csname bibfnamefont\endcsname\relax
  \def\bibfnamefont#1{#1}\fi
\expandafter\ifx\csname citenamefont\endcsname\relax
  \def\citenamefont#1{#1}\fi
\expandafter\ifx\csname url\endcsname\relax
  \def\url#1{\texttt{#1}}\fi
\expandafter\ifx\csname urlprefix\endcsname\relax\def\urlprefix{URL }\fi
\providecommand{\bibinfo}[2]{#2}
\providecommand{\eprint}[2][]{\url{#2}}

\bibitem[{\citenamefont{Maldacena}(1998)}]{Maldacena:1997re}
\bibinfo{author}{\bibfnamefont{J.~M.} \bibnamefont{Maldacena}},
  \bibinfo{journal}{Adv.Theor.Math.Phys.} \textbf{\bibinfo{volume}{2}},
  \bibinfo{pages}{231} (\bibinfo{year}{1998}), \eprint{hep-th/9711200}.

\bibitem[{\citenamefont{Gubser et~al.}(1998)\citenamefont{Gubser, Klebanov, and
  Polyakov}}]{Gubser:1998bc}
\bibinfo{author}{\bibfnamefont{S.}~\bibnamefont{Gubser}},
  \bibinfo{author}{\bibfnamefont{I.~R.} \bibnamefont{Klebanov}},
  \bibnamefont{and} \bibinfo{author}{\bibfnamefont{A.~M.}
  \bibnamefont{Polyakov}}, \bibinfo{journal}{Phys.Lett.}
  \textbf{\bibinfo{volume}{B428}}, \bibinfo{pages}{105} (\bibinfo{year}{1998}),
  \eprint{hep-th/9802109}.

\bibitem[{\citenamefont{Witten}(1998{\natexlab{a}})}]{Witten:1998qj}
\bibinfo{author}{\bibfnamefont{E.}~\bibnamefont{Witten}},
  \bibinfo{journal}{Adv.Theor.Math.Phys.} \textbf{\bibinfo{volume}{2}},
  \bibinfo{pages}{253} (\bibinfo{year}{1998}{\natexlab{a}}),
  \eprint{hep-th/9802150}.

\bibitem[{\citenamefont{Witten}(1998{\natexlab{b}})}]{Witten:1998zw}
\bibinfo{author}{\bibfnamefont{E.}~\bibnamefont{Witten}},
  \bibinfo{journal}{Adv.Theor.Math.Phys.} \textbf{\bibinfo{volume}{2}},
  \bibinfo{pages}{505} (\bibinfo{year}{1998}{\natexlab{b}}),
  \eprint{hep-th/9803131}.

\bibitem[{\citenamefont{Aharony et~al.}(2000)\citenamefont{Aharony, Gubser,
  Maldacena, Ooguri, and Oz}}]{Aharony:1999ti}
\bibinfo{author}{\bibfnamefont{O.}~\bibnamefont{Aharony}},
  \bibinfo{author}{\bibfnamefont{S.~S.} \bibnamefont{Gubser}},
  \bibinfo{author}{\bibfnamefont{J.~M.} \bibnamefont{Maldacena}},
  \bibinfo{author}{\bibfnamefont{H.}~\bibnamefont{Ooguri}}, \bibnamefont{and}
  \bibinfo{author}{\bibfnamefont{Y.}~\bibnamefont{Oz}},
  \bibinfo{journal}{Phys.Rept.} \textbf{\bibinfo{volume}{323}},
  \bibinfo{pages}{183} (\bibinfo{year}{2000}), \eprint{hep-th/9905111}.

\bibitem[{\citenamefont{Shuryak}(2004)}]{Shuryak:2003xe}
\bibinfo{author}{\bibfnamefont{E.}~\bibnamefont{Shuryak}},
  \bibinfo{journal}{Prog.Part.Nucl.Phys.} \textbf{\bibinfo{volume}{53}},
  \bibinfo{pages}{273} (\bibinfo{year}{2004}), \eprint{hep-ph/0312227}.

\bibitem[{\citenamefont{Shuryak}(2005)}]{Shuryak:2004cy}
\bibinfo{author}{\bibfnamefont{E.~V.} \bibnamefont{Shuryak}},
  \bibinfo{journal}{Nucl.Phys.} \textbf{\bibinfo{volume}{A750}},
  \bibinfo{pages}{64} (\bibinfo{year}{2005}), \eprint{hep-ph/0405066}.

\bibitem[{\citenamefont{Adcox et~al.}(2005)}]{Adcox:2004mh}
\bibinfo{author}{\bibfnamefont{K.}~\bibnamefont{Adcox}} \bibnamefont{et~al.}
  (\bibinfo{collaboration}{PHENIX Collaboration}),
  \bibinfo{journal}{Nucl.Phys.} \textbf{\bibinfo{volume}{A757}},
  \bibinfo{pages}{184} (\bibinfo{year}{2005}), \eprint{nucl-ex/0410003}.

\bibitem[{\citenamefont{Adams et~al.}(2005)}]{Adams:2005dq}
\bibinfo{author}{\bibfnamefont{J.}~\bibnamefont{Adams}} \bibnamefont{et~al.}
  (\bibinfo{collaboration}{STAR Collaboration}), \bibinfo{journal}{Nucl.Phys.}
  \textbf{\bibinfo{volume}{A757}}, \bibinfo{pages}{102} (\bibinfo{year}{2005}),
  \eprint{nucl-ex/0501009}.

\bibitem[{\citenamefont{Luzum and Romatschke}(2009)}]{Luzum:2009sb}
\bibinfo{author}{\bibfnamefont{M.}~\bibnamefont{Luzum}} \bibnamefont{and}
  \bibinfo{author}{\bibfnamefont{P.}~\bibnamefont{Romatschke}},
  \bibinfo{journal}{Phys.Rev.Lett.} \textbf{\bibinfo{volume}{103}},
  \bibinfo{pages}{262302} (\bibinfo{year}{2009}), \eprint{0901.4588}.

\bibitem[{\citenamefont{Song et~al.}(2011)\citenamefont{Song, Bass, Heinz,
  Hirano, and Shen}}]{Song:2010mg}
\bibinfo{author}{\bibfnamefont{H.}~\bibnamefont{Song}},
  \bibinfo{author}{\bibfnamefont{S.~A.} \bibnamefont{Bass}},
  \bibinfo{author}{\bibfnamefont{U.}~\bibnamefont{Heinz}},
  \bibinfo{author}{\bibfnamefont{T.}~\bibnamefont{Hirano}}, \bibnamefont{and}
  \bibinfo{author}{\bibfnamefont{C.}~\bibnamefont{Shen}},
  \bibinfo{journal}{Phys.Rev.Lett.} \textbf{\bibinfo{volume}{106}},
  \bibinfo{pages}{192301} (\bibinfo{year}{2011}), \eprint{1011.2783}.

\bibitem[{\citenamefont{Qiu et~al.}(2012)\citenamefont{Qiu, Shen, and
  Heinz}}]{Qiu:2011hf}
\bibinfo{author}{\bibfnamefont{Z.}~\bibnamefont{Qiu}},
  \bibinfo{author}{\bibfnamefont{C.}~\bibnamefont{Shen}}, \bibnamefont{and}
  \bibinfo{author}{\bibfnamefont{U.}~\bibnamefont{Heinz}},
  \bibinfo{journal}{Phys.Lett.} \textbf{\bibinfo{volume}{B707}},
  \bibinfo{pages}{151} (\bibinfo{year}{2012}), \eprint{1110.3033}.

\bibitem[{\citenamefont{Schenke et~al.}(2012)\citenamefont{Schenke, Jeon, and
  Gale}}]{Schenke:2011bn}
\bibinfo{author}{\bibfnamefont{B.}~\bibnamefont{Schenke}},
  \bibinfo{author}{\bibfnamefont{S.}~\bibnamefont{Jeon}}, \bibnamefont{and}
  \bibinfo{author}{\bibfnamefont{C.}~\bibnamefont{Gale}},
  \bibinfo{journal}{Phys.Rev.} \textbf{\bibinfo{volume}{C85}},
  \bibinfo{pages}{024901} (\bibinfo{year}{2012}), \eprint{1109.6289}.

\bibitem[{\citenamefont{Gale et~al.}(2012)\citenamefont{Gale, Jeon, Schenke,
  Tribedy, and Venugopalan}}]{Gale:2012rq}
\bibinfo{author}{\bibfnamefont{C.}~\bibnamefont{Gale}},
  \bibinfo{author}{\bibfnamefont{S.}~\bibnamefont{Jeon}},
  \bibinfo{author}{\bibfnamefont{B.}~\bibnamefont{Schenke}},
  \bibinfo{author}{\bibfnamefont{P.}~\bibnamefont{Tribedy}}, \bibnamefont{and}
  \bibinfo{author}{\bibfnamefont{R.}~\bibnamefont{Venugopalan}}
  (\bibinfo{year}{2012}), \eprint{1209.6330}.

\bibitem[{\citenamefont{Policastro et~al.}(2001)\citenamefont{Policastro, Son,
  and Starinets}}]{Policastro:2001yc}
\bibinfo{author}{\bibfnamefont{G.}~\bibnamefont{Policastro}},
  \bibinfo{author}{\bibfnamefont{D.}~\bibnamefont{Son}}, \bibnamefont{and}
  \bibinfo{author}{\bibfnamefont{A.}~\bibnamefont{Starinets}},
  \bibinfo{journal}{Phys.Rev.Lett.} \textbf{\bibinfo{volume}{87}},
  \bibinfo{pages}{081601} (\bibinfo{year}{2001}), \eprint{hep-th/0104066}.

\bibitem[{\citenamefont{Kovtun et~al.}(2005)\citenamefont{Kovtun, Son, and
  Starinets}}]{Kovtun:2004de}
\bibinfo{author}{\bibfnamefont{P.}~\bibnamefont{Kovtun}},
  \bibinfo{author}{\bibfnamefont{D.}~\bibnamefont{Son}}, \bibnamefont{and}
  \bibinfo{author}{\bibfnamefont{A.}~\bibnamefont{Starinets}},
  \bibinfo{journal}{Phys.Rev.Lett.} \textbf{\bibinfo{volume}{94}},
  \bibinfo{pages}{111601} (\bibinfo{year}{2005}), \eprint{hep-th/0405231}.

\bibitem[{\citenamefont{Florkowski}(2008)}]{Florkowski:2008ag}
\bibinfo{author}{\bibfnamefont{W.}~\bibnamefont{Florkowski}},
  \bibinfo{journal}{Phys.Lett.} \textbf{\bibinfo{volume}{B668}},
  \bibinfo{pages}{32} (\bibinfo{year}{2008}), \eprint{0806.2268}.

\bibitem[{\citenamefont{Florkowski and Ryblewski}(2009)}]{Florkowski:2009sw}
\bibinfo{author}{\bibfnamefont{W.}~\bibnamefont{Florkowski}} \bibnamefont{and}
  \bibinfo{author}{\bibfnamefont{R.}~\bibnamefont{Ryblewski}},
  \bibinfo{journal}{Acta Phys.Polon.} \textbf{\bibinfo{volume}{B40}},
  \bibinfo{pages}{2843} (\bibinfo{year}{2009}), \eprint{0901.4653}.

\bibitem[{\citenamefont{Ryblewski and Florkowski}(2010)}]{Ryblewski:2010tn}
\bibinfo{author}{\bibfnamefont{R.}~\bibnamefont{Ryblewski}} \bibnamefont{and}
  \bibinfo{author}{\bibfnamefont{W.}~\bibnamefont{Florkowski}},
  \bibinfo{journal}{Phys.Rev.} \textbf{\bibinfo{volume}{C82}},
  \bibinfo{pages}{024903} (\bibinfo{year}{2010}), \eprint{1004.1594}.

\bibitem[{\citenamefont{Florkowski and Ryblewski}(2011)}]{Florkowski:2010cf}
\bibinfo{author}{\bibfnamefont{W.}~\bibnamefont{Florkowski}} \bibnamefont{and}
  \bibinfo{author}{\bibfnamefont{R.}~\bibnamefont{Ryblewski}},
  \bibinfo{journal}{Phys.Rev.} \textbf{\bibinfo{volume}{C83}},
  \bibinfo{pages}{034907} (\bibinfo{year}{2011}), \eprint{1007.0130}.

\bibitem[{\citenamefont{Martinez and Strickland}(2010)}]{Martinez:2010sc}
\bibinfo{author}{\bibfnamefont{M.}~\bibnamefont{Martinez}} \bibnamefont{and}
  \bibinfo{author}{\bibfnamefont{M.}~\bibnamefont{Strickland}},
  \bibinfo{journal}{Nucl.Phys.} \textbf{\bibinfo{volume}{A848}},
  \bibinfo{pages}{183} (\bibinfo{year}{2010}), \eprint{1007.0889}.

\bibitem[{\citenamefont{Martinez and Strickland}(2011)}]{Martinez:2010sd}
\bibinfo{author}{\bibfnamefont{M.}~\bibnamefont{Martinez}} \bibnamefont{and}
  \bibinfo{author}{\bibfnamefont{M.}~\bibnamefont{Strickland}},
  \bibinfo{journal}{Nucl.Phys.} \textbf{\bibinfo{volume}{A856}},
  \bibinfo{pages}{68} (\bibinfo{year}{2011}), \eprint{1011.3056}.

\bibitem[{\citenamefont{Ryblewski and
  Florkowski}(2011{\natexlab{a}})}]{Ryblewski:2010ch}
\bibinfo{author}{\bibfnamefont{R.}~\bibnamefont{Ryblewski}} \bibnamefont{and}
  \bibinfo{author}{\bibfnamefont{W.}~\bibnamefont{Florkowski}},
  \bibinfo{journal}{Acta Phys.Polon.} \textbf{\bibinfo{volume}{B42}},
  \bibinfo{pages}{115} (\bibinfo{year}{2011}{\natexlab{a}}),
  \eprint{1011.6213}.

\bibitem[{\citenamefont{Ryblewski and
  Florkowski}(2011{\natexlab{b}})}]{Ryblewski:2011aq}
\bibinfo{author}{\bibfnamefont{R.}~\bibnamefont{Ryblewski}} \bibnamefont{and}
  \bibinfo{author}{\bibfnamefont{W.}~\bibnamefont{Florkowski}},
  \bibinfo{journal}{Eur.Phys.J.} \textbf{\bibinfo{volume}{C71}},
  \bibinfo{pages}{1761} (\bibinfo{year}{2011}{\natexlab{b}}),
  \eprint{1103.1260}.

\bibitem[{\citenamefont{Strickland}(2011)}]{Strickland:2011mw}
\bibinfo{author}{\bibfnamefont{M.}~\bibnamefont{Strickland}},
  \bibinfo{journal}{Phys.Rev.Lett.} \textbf{\bibinfo{volume}{107}},
  \bibinfo{pages}{132301} (\bibinfo{year}{2011}), \eprint{1106.2571}.

\bibitem[{\citenamefont{Strickland and Bazow}(2012)}]{Strickland:2011aa}
\bibinfo{author}{\bibfnamefont{M.}~\bibnamefont{Strickland}} \bibnamefont{and}
  \bibinfo{author}{\bibfnamefont{D.}~\bibnamefont{Bazow}},
  \bibinfo{journal}{Nucl.Phys.} \textbf{\bibinfo{volume}{A879}},
  \bibinfo{pages}{25} (\bibinfo{year}{2012}), \eprint{1112.2761}.

\bibitem[{\citenamefont{Martinez et~al.}(2012)\citenamefont{Martinez,
  Ryblewski, and Strickland}}]{Martinez:2012tu}
\bibinfo{author}{\bibfnamefont{M.}~\bibnamefont{Martinez}},
  \bibinfo{author}{\bibfnamefont{R.}~\bibnamefont{Ryblewski}},
  \bibnamefont{and}
  \bibinfo{author}{\bibfnamefont{M.}~\bibnamefont{Strickland}},
  \bibinfo{journal}{Phys.Rev.} \textbf{\bibinfo{volume}{C85}},
  \bibinfo{pages}{064913} (\bibinfo{year}{2012}), \eprint{1204.1473}.

\bibitem[{\citenamefont{Florkowski et~al.}(2012)\citenamefont{Florkowski, Maj,
  Ryblewski, and Strickland}}]{Florkowski:2012as}
\bibinfo{author}{\bibfnamefont{W.}~\bibnamefont{Florkowski}},
  \bibinfo{author}{\bibfnamefont{R.}~\bibnamefont{Maj}},
  \bibinfo{author}{\bibfnamefont{R.}~\bibnamefont{Ryblewski}},
  \bibnamefont{and}
  \bibinfo{author}{\bibfnamefont{M.}~\bibnamefont{Strickland}}
  (\bibinfo{year}{2012}), \eprint{1209.3671}.

\bibitem[{\citenamefont{Heller et~al.}(2011)\citenamefont{Heller, Janik, and
  Witaszczyk}}]{Heller:2011ju}
\bibinfo{author}{\bibfnamefont{M.~P.} \bibnamefont{Heller}},
  \bibinfo{author}{\bibfnamefont{R.~A.} \bibnamefont{Janik}}, \bibnamefont{and}
  \bibinfo{author}{\bibfnamefont{P.}~\bibnamefont{Witaszczyk}}
  (\bibinfo{year}{2011}), \eprint{1103.3452}.

\bibitem[{\citenamefont{Heller et~al.}(2012)\citenamefont{Heller, Janik, and
  Witaszczyk}}]{Heller:2012je}
\bibinfo{author}{\bibfnamefont{M.~P.} \bibnamefont{Heller}},
  \bibinfo{author}{\bibfnamefont{R.~A.} \bibnamefont{Janik}}, \bibnamefont{and}
  \bibinfo{author}{\bibfnamefont{P.}~\bibnamefont{Witaszczyk}}
  (\bibinfo{year}{2012}), \eprint{1203.0755}.

\bibitem[{\citenamefont{Janik and Witaszczyk}(2008)}]{Janik:2008tc}
\bibinfo{author}{\bibfnamefont{R.~A.} \bibnamefont{Janik}} \bibnamefont{and}
  \bibinfo{author}{\bibfnamefont{P.}~\bibnamefont{Witaszczyk}},
  \bibinfo{journal}{JHEP} \textbf{\bibinfo{volume}{0809}}, \bibinfo{pages}{026}
  (\bibinfo{year}{2008}), \eprint{0806.2141}.

\bibitem[{\citenamefont{Mateos and
  Trancanelli}(2011{\natexlab{a}})}]{Mateos:2011ix}
\bibinfo{author}{\bibfnamefont{D.}~\bibnamefont{Mateos}} \bibnamefont{and}
  \bibinfo{author}{\bibfnamefont{D.}~\bibnamefont{Trancanelli}},
  \bibinfo{journal}{Phys.Rev.Lett.} \textbf{\bibinfo{volume}{107}},
  \bibinfo{pages}{101601} (\bibinfo{year}{2011}{\natexlab{a}}),
  \eprint{1105.3472}.

\bibitem[{\citenamefont{Mateos and
  Trancanelli}(2011{\natexlab{b}})}]{Mateos:2011tv}
\bibinfo{author}{\bibfnamefont{D.}~\bibnamefont{Mateos}} \bibnamefont{and}
  \bibinfo{author}{\bibfnamefont{D.}~\bibnamefont{Trancanelli}},
  \bibinfo{journal}{JHEP} \textbf{\bibinfo{volume}{1107}}, \bibinfo{pages}{054}
  (\bibinfo{year}{2011}{\natexlab{b}}), \eprint{1106.1637}.

\bibitem[{\citenamefont{Gahramanov et~al.}(2012)\citenamefont{Gahramanov,
  Kalaydzhyan, and Kirsch}}]{Gahramanov:2012wz}
\bibinfo{author}{\bibfnamefont{I.}~\bibnamefont{Gahramanov}},
  \bibinfo{author}{\bibfnamefont{T.}~\bibnamefont{Kalaydzhyan}},
  \bibnamefont{and} \bibinfo{author}{\bibfnamefont{I.}~\bibnamefont{Kirsch}},
  \bibinfo{journal}{Phys.Rev.} \textbf{\bibinfo{volume}{D85}},
  \bibinfo{pages}{126013} (\bibinfo{year}{2012}), \eprint{1203.4259}.

\bibitem[{\citenamefont{Gubser}(2007)}]{Gubser:2006qh}
\bibinfo{author}{\bibfnamefont{S.~S.} \bibnamefont{Gubser}},
  \bibinfo{journal}{Phys. Rev.} \textbf{\bibinfo{volume}{D76}},
  \bibinfo{pages}{126003} (\bibinfo{year}{2007}), \eprint{hep-th/0611272}.

\bibitem[{\citenamefont{Herzog et~al.}(2006)\citenamefont{Herzog, Karch,
  Kovtun, Kozcaz, and Yaffe}}]{Herzog:2006gh}
\bibinfo{author}{\bibfnamefont{C.}~\bibnamefont{Herzog}},
  \bibinfo{author}{\bibfnamefont{A.}~\bibnamefont{Karch}},
  \bibinfo{author}{\bibfnamefont{P.}~\bibnamefont{Kovtun}},
  \bibinfo{author}{\bibfnamefont{C.}~\bibnamefont{Kozcaz}}, \bibnamefont{and}
  \bibinfo{author}{\bibfnamefont{L.}~\bibnamefont{Yaffe}},
  \bibinfo{journal}{JHEP} \textbf{\bibinfo{volume}{0607}}, \bibinfo{pages}{013}
  (\bibinfo{year}{2006}), \eprint{hep-th/0605158}.

\bibitem[{\citenamefont{Liu et~al.}(2006)\citenamefont{Liu, Rajagopal, and
  Wiedemann}}]{Liu:2006ug}
\bibinfo{author}{\bibfnamefont{H.}~\bibnamefont{Liu}},
  \bibinfo{author}{\bibfnamefont{K.}~\bibnamefont{Rajagopal}},
  \bibnamefont{and} \bibinfo{author}{\bibfnamefont{U.~A.}
  \bibnamefont{Wiedemann}}, \bibinfo{journal}{Phys.Rev.Lett.}
  \textbf{\bibinfo{volume}{97}}, \bibinfo{pages}{182301}
  (\bibinfo{year}{2006}), \eprint{hep-ph/0605178}.

\bibitem[{\citenamefont{Liu et~al.}(2007)\citenamefont{Liu, Rajagopal, and
  Wiedemann}}]{Liu:2006he}
\bibinfo{author}{\bibfnamefont{H.}~\bibnamefont{Liu}},
  \bibinfo{author}{\bibfnamefont{K.}~\bibnamefont{Rajagopal}},
  \bibnamefont{and} \bibinfo{author}{\bibfnamefont{U.~A.}
  \bibnamefont{Wiedemann}}, \bibinfo{journal}{JHEP}
  \textbf{\bibinfo{volume}{0703}}, \bibinfo{pages}{066} (\bibinfo{year}{2007}),
  \eprint{hep-ph/0612168}.

\bibitem[{\citenamefont{Giataganas}(2012)}]{Giataganas:2012zy}
\bibinfo{author}{\bibfnamefont{D.}~\bibnamefont{Giataganas}},
  \bibinfo{journal}{JHEP} \textbf{\bibinfo{volume}{1207}}, \bibinfo{pages}{031}
  (\bibinfo{year}{2012}), \eprint{1202.4436}.

\bibitem[{\citenamefont{Rebhan and Steineder}(2012)}]{Rebhan:2012bw}
\bibinfo{author}{\bibfnamefont{A.}~\bibnamefont{Rebhan}} \bibnamefont{and}
  \bibinfo{author}{\bibfnamefont{D.}~\bibnamefont{Steineder}},
  \bibinfo{journal}{JHEP} \textbf{\bibinfo{volume}{1208}}, \bibinfo{pages}{020}
  (\bibinfo{year}{2012}), \eprint{1205.4684}.

\bibitem[{\citenamefont{Chernicoff
  et~al.}(2012{\natexlab{a}})\citenamefont{Chernicoff, Fernandez, Mateos, and
  Trancanelli}}]{Chernicoff:2012gu}
\bibinfo{author}{\bibfnamefont{M.}~\bibnamefont{Chernicoff}},
  \bibinfo{author}{\bibfnamefont{D.}~\bibnamefont{Fernandez}},
  \bibinfo{author}{\bibfnamefont{D.}~\bibnamefont{Mateos}}, \bibnamefont{and}
  \bibinfo{author}{\bibfnamefont{D.}~\bibnamefont{Trancanelli}},
  \bibinfo{journal}{JHEP} \textbf{\bibinfo{volume}{1208}}, \bibinfo{pages}{041}
  (\bibinfo{year}{2012}{\natexlab{a}}), \eprint{1203.0561}.

\bibitem[{\citenamefont{Chernicoff
  et~al.}(2012{\natexlab{b}})\citenamefont{Chernicoff, Fernandez, Mateos, and
  Trancanelli}}]{Chernicoff:2012iq}
\bibinfo{author}{\bibfnamefont{M.}~\bibnamefont{Chernicoff}},
  \bibinfo{author}{\bibfnamefont{D.}~\bibnamefont{Fernandez}},
  \bibinfo{author}{\bibfnamefont{D.}~\bibnamefont{Mateos}}, \bibnamefont{and}
  \bibinfo{author}{\bibfnamefont{D.}~\bibnamefont{Trancanelli}},
  \bibinfo{journal}{JHEP} \textbf{\bibinfo{volume}{1208}}, \bibinfo{pages}{100}
  (\bibinfo{year}{2012}{\natexlab{b}}), \eprint{1202.3696}.

\bibitem[{\citenamefont{Fadafan and Soltanpanahi}(2012)}]{Fadafan:2012qu}
\bibinfo{author}{\bibfnamefont{K.~B.} \bibnamefont{Fadafan}} \bibnamefont{and}
  \bibinfo{author}{\bibfnamefont{H.}~\bibnamefont{Soltanpanahi}}
  (\bibinfo{year}{2012}), \eprint{1206.2271}.

\bibitem[{\citenamefont{Chernicoff
  et~al.}(2012{\natexlab{c}})\citenamefont{Chernicoff, Fernandez, Mateos, and
  Trancanelli}}]{Chernicoff:2012bu}
\bibinfo{author}{\bibfnamefont{M.}~\bibnamefont{Chernicoff}},
  \bibinfo{author}{\bibfnamefont{D.}~\bibnamefont{Fernandez}},
  \bibinfo{author}{\bibfnamefont{D.}~\bibnamefont{Mateos}}, \bibnamefont{and}
  \bibinfo{author}{\bibfnamefont{D.}~\bibnamefont{Trancanelli}}
  (\bibinfo{year}{2012}{\natexlab{c}}), \eprint{1208.2672}.

\bibitem[{\citenamefont{Gubser et~al.}(2008)\citenamefont{Gubser, Gulotta,
  Pufu, and Rocha}}]{Gubser:2008as}
\bibinfo{author}{\bibfnamefont{S.~S.} \bibnamefont{Gubser}},
  \bibinfo{author}{\bibfnamefont{D.~R.} \bibnamefont{Gulotta}},
  \bibinfo{author}{\bibfnamefont{S.~S.} \bibnamefont{Pufu}}, \bibnamefont{and}
  \bibinfo{author}{\bibfnamefont{F.~D.} \bibnamefont{Rocha}},
  \bibinfo{journal}{JHEP} \textbf{\bibinfo{volume}{10}}, \bibinfo{pages}{052}
  (\bibinfo{year}{2008}), \eprint{0803.1470}.

\bibitem[{\citenamefont{Chesler et~al.}(2009)\citenamefont{Chesler, Jensen,
  Karch, and Yaffe}}]{Chesler:2008uy}
\bibinfo{author}{\bibfnamefont{P.~M.} \bibnamefont{Chesler}},
  \bibinfo{author}{\bibfnamefont{K.}~\bibnamefont{Jensen}},
  \bibinfo{author}{\bibfnamefont{A.}~\bibnamefont{Karch}}, \bibnamefont{and}
  \bibinfo{author}{\bibfnamefont{L.~G.} \bibnamefont{Yaffe}},
  \bibinfo{journal}{Phys. Rev.} \textbf{\bibinfo{volume}{D79}},
  \bibinfo{pages}{125015} (\bibinfo{year}{2009}), \eprint{0810.1985}.

\bibitem[{\citenamefont{Arnold and Vaman}(2010)}]{Arnold:2010ir}
\bibinfo{author}{\bibfnamefont{P.}~\bibnamefont{Arnold}} \bibnamefont{and}
  \bibinfo{author}{\bibfnamefont{D.}~\bibnamefont{Vaman}},
  \bibinfo{journal}{JHEP} \textbf{\bibinfo{volume}{10}}, \bibinfo{pages}{099}
  (\bibinfo{year}{2010}), \eprint{1008.4023}.

\bibitem[{\citenamefont{Arnold and Vaman}(2011)}]{Arnold:2011qi}
\bibinfo{author}{\bibfnamefont{P.}~\bibnamefont{Arnold}} \bibnamefont{and}
  \bibinfo{author}{\bibfnamefont{D.}~\bibnamefont{Vaman}},
  \bibinfo{journal}{JHEP} \textbf{\bibinfo{volume}{1104}}, \bibinfo{pages}{027}
  (\bibinfo{year}{2011}), \eprint{1101.2689}.

\bibitem[{\citenamefont{Arnold et~al.}(2012)\citenamefont{Arnold, Szepietowski,
  and Vaman}}]{Arnold:2012uc}
\bibinfo{author}{\bibfnamefont{P.}~\bibnamefont{Arnold}},
  \bibinfo{author}{\bibfnamefont{P.}~\bibnamefont{Szepietowski}},
  \bibnamefont{and} \bibinfo{author}{\bibfnamefont{D.}~\bibnamefont{Vaman}}
  (\bibinfo{year}{2012}), \eprint{1203.6658}.

\bibitem[{\citenamefont{Romatschke and Strickland}(2003)}]{Romatschke:2003ms}
\bibinfo{author}{\bibfnamefont{P.}~\bibnamefont{Romatschke}} \bibnamefont{and}
  \bibinfo{author}{\bibfnamefont{M.}~\bibnamefont{Strickland}},
  \bibinfo{journal}{Phys.Rev.} \textbf{\bibinfo{volume}{D68}},
  \bibinfo{pages}{036004} (\bibinfo{year}{2003}), \eprint{hep-ph/0304092}.

\bibitem[{\citenamefont{Romatschke and Strickland}(2004)}]{Romatschke:2004jh}
\bibinfo{author}{\bibfnamefont{P.}~\bibnamefont{Romatschke}} \bibnamefont{and}
  \bibinfo{author}{\bibfnamefont{M.}~\bibnamefont{Strickland}},
  \bibinfo{journal}{Phys.Rev.} \textbf{\bibinfo{volume}{D70}},
  \bibinfo{pages}{116006} (\bibinfo{year}{2004}), \eprint{hep-ph/0406188}.

\bibitem[{\citenamefont{Hatta et~al.}(2008)\citenamefont{Hatta, Iancu, and
  Mueller}}]{Hatta:2008tx}
\bibinfo{author}{\bibfnamefont{Y.}~\bibnamefont{Hatta}},
  \bibinfo{author}{\bibfnamefont{E.}~\bibnamefont{Iancu}}, \bibnamefont{and}
  \bibinfo{author}{\bibfnamefont{A.~H.} \bibnamefont{Mueller}},
  \bibinfo{journal}{JHEP} \textbf{\bibinfo{volume}{05}}, \bibinfo{pages}{037}
  (\bibinfo{year}{2008}), \eprint{0803.2481}.

\bibitem[{\citenamefont{Panero}(2009)}]{Panero:2009tv}
\bibinfo{author}{\bibfnamefont{M.}~\bibnamefont{Panero}},
  \bibinfo{journal}{Phys.Rev.Lett.} \textbf{\bibinfo{volume}{103}},
  \bibinfo{pages}{232001} (\bibinfo{year}{2009}), \eprint{0907.3719}.

\bibitem[{\citenamefont{Baier}(2003)}]{Baier:2002tc}
\bibinfo{author}{\bibfnamefont{R.}~\bibnamefont{Baier}},
  \bibinfo{journal}{Nucl.Phys.} \textbf{\bibinfo{volume}{A715}},
  \bibinfo{pages}{209} (\bibinfo{year}{2003}), \eprint{hep-ph/0209038}.

\end{thebibliography}

\end{document}